\documentclass{article}
\usepackage{PRIMEarxiv}

\usepackage[utf8]{inputenc} % allow utf-8 input
\usepackage[T1]{fontenc}    % use 8-bit T1 fonts
\usepackage{hyperref}       % hyperlinks
\usepackage{url}            % simple URL typesetting
\usepackage{booktabs}       % professional-quality tables
\usepackage{amsfonts}       % blackboard math symbols
\usepackage{nicefrac}       % compact symbols for 1/2, etc.
\usepackage{microtype}      % micro typography
\usepackage{lipsum}
\usepackage{fancyhdr}       % header
\usepackage{graphicx}       % graphics
\graphicspath{{media/}}     % organize your images and other figures
\usepackage{multirow}
\usepackage{tikz}
\usetikzlibrary{shapes, arrows.meta, positioning}
\usepackage{algorithm}
\usepackage{algorithmic}
\usepackage{amsmath} % For mathematical symbols
\usepackage{amssymb} % For additional math symbols

\title{A Domain Adaptation Model for Carotid Ultrasound: Image Harmonization, Noise Reduction, and Impact on Cardiovascular Risk Markers}

\author{
  Mohd Usama$^{\mathrm{\ast}}$ \\
  \texttt{mohd.usama@umu.se} 
   \And
  Emma Nyman$^{\mathrm{\dag}}$ \\
  \texttt{emma.nyman@umu.se} 
     \And
  Ulf Näslund$^{\mathrm{\dag}}$ \\
  \texttt{ulf.naslund@umu.se} 
     \And
  Christer Grönlund$^{\mathrm{\ast}}$ \\
  \texttt{christer.gronlund@umu.se}
    \And
  \\
  \\
  \textit{$^{\mathrm{\ast}}$Department of Diagnostics and Intervention, and Biomedical Engineering, Umea University, Umea, Sweden}\\
  \textit{$^{\mathrm{\dag}}$Department of Public Health and Clinical Medicine, Umea University, Umea, Sweden}\\\\
  \scshape A preprint - \today
}

\begin{document}
\maketitle

\begin{abstract}
Deep learning has been used extensively for medical image analysis applications, assuming the training and test data adhere to the same probability distributions. However, a common challenge arises when dealing with medical images generated by different systems or even the same system with varying parameter settings. Such images often contain diverse textures and noise patterns, violating the assumption. Consequently, models trained on data from one machine or setting usually struggle to perform effectively on data from another. To address this issue in ultrasound images, we proposed a Generative Adversarial Network (GAN) based model in this paper. We formulated image harmonization and denoising tasks as an image-to-image translation task, wherein we modified the texture pattern and reduced noise in Carotid ultrasound images while keeping the image content (the anatomy) unchanged. 
The performance was evaluated using feature distribution and pixel-space similarity metrics. In addition, blood-to-tissue contrast and influence on computed risk markers (Gray scale median, GSM) were evaluated.
The results showed that domain adaptation was achieved in both tasks (histogram correlation 0.920 and 0.844), as compared to no adaptation (0.890 and 0.707), and that the anatomy of the images was retained (structure similarity index measure of the arterial wall 0.71 and 0.80). In addition, the image noise level (contrast) did not change in the image harmonization task (-34.1 vs 35.2 dB) but was improved in the noise reduction task (-23.5 vs -46.7 dB). 
The model outperformed the CycleGAN in both tasks. Finally, the risk marker GSM increased by 7.6 (p<0.001) in task 1 but not in task 2.
We conclude that domain translation models are powerful tools for ultrasound image improvement while retaining the underlying anatomy but that downstream calculations of risk markers may be affected.
\end{abstract}

% keywords can be removed

\section{Introduction}
Deep learning has been used extensively for medical image analysis applications, assuming the training and test data adhere to the same probability distributions. However, a common challenge arises when dealing with medical images generated by different systems or even the same system with varying parameter settings. Such images often contain diverse textures and noise patterns, violating the assumption. Consequently, models trained on data from one machine or setting usually struggle to perform effectively on data from another.

Deep learning for automatic analysis of ultrasound images has attracted much attention for multiple clinical applications, such as atherosclerosis risk assessment \cite{ward2020machine, fan2023comparing}, artery plaque detection and segmentation \cite{jashari2016carotid, chen2023ultrasound}, breast cancer detection and segmentation \cite{mubashar2022r2u++}, etc. However, images from different ultrasound systems often vary in many aspects due to different imaging parameter settings, algorithm design, hardware components, etc. Therefore, the texture pattern of images acquired by different ultrasound systems may not belong to the same feature population distribution. For example, images taken from different ultrasound machines or different settings usually differ in texture pattern and noise. In addition, the images’ texture and noise can vary depending on the composition and distribution of tissues of the imaged subject. This may cause generalizability problems when using pre-trained models for new images from different systems or on images from patients with varying noise distributions.

To overcome these limitations, several studies have leveraged Generative Adversarial Networks (GANs) \cite{goodfellow2014generative} in ultrasound imaging for domain adaptation and denoising individually. These networks, known for their ability to generate realistic images, have shown promise in various areas of medical imaging, such as denoising \cite{wang2020deep}, domain adaptation \cite{guan2021domain}, disease translation \cite{ali2023translation}, synthetic image generation \cite{fujioka2021virtual}, segmentation\cite{mubashar2022r2u++}, and classification \cite{christodoulou2003texture}. Several researchers have introduced different methods and techniques in ultrasound image processing to enhance these networks' effectiveness and efficiency \cite{litjens2017survey}. They have explored the potential of GANs, which has resulted in a surge of publications focusing on quality adaptation and denoising \cite{zhang2019noise}. Zhou et al. \cite{zhou2019image} proposed a two-stage GAN for translating low-quality ultrasound images into high-quality images, which are usually degraded due to the limited size of portable ultrasound devices. Xia et al. \cite{xia2022multilevel} propose a Multilevel Structure-Preserved Generative Adversarial Network (MSP-GAN) for domain adaptation of ultrasound images while maintaining image intravascular structures. The CycleGAN network \cite{zhu2017unpaired} has been widely used for different applications in domain-to-domain translation tasks that do not require paired image data. Its application spans various modalities of medical images, encompassing ultrasound, MRI and CT \cite{teng2020interactive, zhang2018translating}. Moreover, CycleGAN architecture has shown promising histopathology results in assisting diagnoses more precisely by standardizing microscopy staining \cite{long2017deep}. Additionally, several other works used CycleGAN for image quality adaptation and denoising, such as Athreya et al. \cite{athreya2023ultrasound}, who proposed a CycleGAN-based model with perceptual loss for quality enhancement of ultrasound images. Huang et al. \cite{huang2022stability} proposed a stability-enhanced CycleGAN-based model to harmonize ultrasound images from various medical centres through domain transformation with good preservation of image details. Tan et al.\cite{tan2022selective} proposed an unsupervised model using CycleGAN with selective kernel networks for adapting select features, patchGAN discriminator and perceptual loss in LDCT imaging to generate CT.  Tang et al. \cite{tang2019unpaired} proposed a CycleGAN-based denoising network for low-dose unpaired CT images with prior image information. Similar work is also done by Zhao et al.\cite{zhao2023dual}, who proposed a CycleGAN-based dual-scale similarity-guided model.

Image harmonization in carotid ultrasound images is crucial for standardizing measurements and reducing variability across different scanners and imaging conditions. Various methods have been developed to address this issue. For instance, a study by Elatrozy et al. \cite{elatrozy1998effect} demonstrated the standardization of plaque echodensity in carotid ultrasound images, allowing for comparable measurements. Furthermore, Kakkos et al. \cite{kakkos2006effect} highlighted the importance of standardizing resolution settings during ultrasound scanning to minimize variability in texture analysis of carotid plaques, emphasizing the need for resolution normalization algorithms. Additionally, model-based unsupervised domain adaptation has been successfully applied in vessel segmentation tasks, where neural networks are initially trained on simulated ultrasound images and then fine-tuned on in-vivo data, resulting in significant performance improvements without the need for extensive manual labelling \cite{sharifzadeh2021ultrasound}. These approaches demonstrate the potential of leveraging advanced techniques to enhance the generalizability and accuracy of carotid ultrasound image analysis through domain adaptation and harmonization strategies. Moreover, Cheung et al. \cite{cheung2015attenuation} developed an algorithm to correct attenuation artifacts and normalize intensity in contrast-enhanced ultrasound images of carotid arteries, enhancing quantification and reducing variability, and Loizou et al. \cite{loizou2006quality} focus on evaluating the quality of ultrasound images in the carotid artery after applying normalization and speckle reduction filtering techniques. Image quality assessment is crucial for tasks like assessing atherosclerotic disease and telemedicine image transfer. Noise reduction in ultrasound images plays another crucial role in enhancing the quality of the images. Research has shown that GANs, such as the wavelet-based GAN proposed in the studies \cite{kemna2023reduced, khor2022ultrasound}, can effectively suppress speckle noise while preserving image features, leading to high-quality image reconstruction. Additionally, the use of GANs in ultrasound image processing, as demonstrated in various studies \cite{asiedu2022generative, seoni2022ultrasound, zhang2022ultrasound}, has shown promising results in improving image quality. By leveraging GANs, particularly customized models like conditional GANs and wavelet-based GANs, researchers have been able to address the challenges of noise reduction in carotid ultrasound images, ensuring accurate diagnostic information is preserved for clinical applications.

Taken together, previous work shows that, among others, image quality adaptation and denoising are important challenges in carotid ultrasound images, and separate models have been developed to solve these tasks individually \cite{wang2020deep}. However, this requires building different models for each task, including data collection from each machine with similar parameter settings to train the models, which is time-consuming and tedious, requiring separate resources and computation power, which is expensive \cite{long2017deep, meng2022dual, zhu2022transfer}.

This paper proposes a new GAN-based model that can address image harmonization and image denoising tasks with the same architecture. 
The proposed model contains one generator and two discriminators, and model training is regularized with content, noise and adversarial losses. The generator aims to take a source image and generate an image with features of the target domain. The first discriminator with content loss ensures that the generated image’s content (anatomy) is preserved, and the second discriminator with noise loss enforces the noise features in the image.

The main contributions of this paper are:
\begin{itemize}
\item We propose a GAN-based model to address the problem of denoising and domain adaptation of ultrasound images through an unpaired image-to-image translation task. The model is trained to transfer target domain image features into the source domain image without changing the underlying anatomical content of the image.
\item We introduce the noise loss, which employs the Wasserstein distance to measure the dissimilarity between the stylistic features derived from the generator's early layer and the style of target domain images.
\item We fine-tune the generator architecture from Jun-Yan Zhu et al. \cite{zhu2017unpaired} by adding two extra blocks (one convolution block and one deconvolution block) to manage the feature map extraction from the first three layers.
\item We evaluate the performance of the domain adaptation task and preservation of anatomical content. In addition, we evaluate the domain adaptation’s impact on clinical risk markers. To our knowledge, this is the first study to do so.
\end{itemize}

The rest of the paper is organized as follows. Section II explains the method. Section III describes the results, and section IV discusses the findings. The last section concludes the paper.

\section{Methods}
This paper presents an unpaired image-to-image translation model such that an input image with one feature distribution can be converted to an output image with another feature distribution while preserving the underlying anatomical content of the images. We target the problems of image feature harmonization and noise reduction. Generally, paired images from the corresponding domains (different systems, different noise levels) are often unavailable. Therefore, we chose to solve this as an unpaired image-to-image translation task. In this work, we consider the underlying anatomical information of the images as content and noise pattern as a style and formulate the task as a style transfer task.

\subsection{GAN-based Model}
\begin{figure*}
\centering
{\includegraphics[scale=0.95]{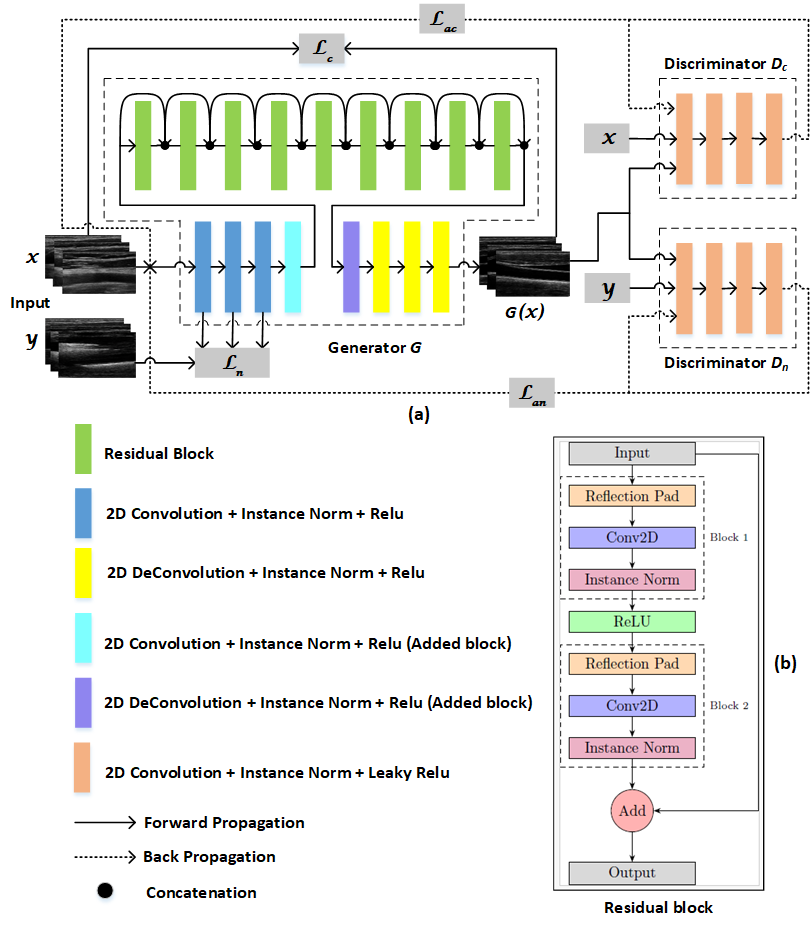}}
\caption{a) Proposed model. $G$ denotes the generator network and two discriminators, $D_c$ and $D_n$, trained by adversarial losses $\mathcal{L}_{ac}$ and $\mathcal{L}_{an}$, respectively. $x \sim S$ denotes the image set from the source domain. $y \sim T$ represents the image set from the target domain. $G(x)$ denotes the image set generated by $G$. $L_c$ indicates the content loss calculated from the feature difference of the source and $G(x)$ image. $L_n$ represents the noise loss calculated from the feature difference between the target image and the early layers of the generator. b) a residual block.}
\label{Figure1}
\end{figure*}

To solve the aforementioned problem, we propose a GAN-based model consisting of one generator and two discriminators. We formulate the task as an unpaired image-to-image translation task and incorporate a noise loss. This loss is determined using the Wasserstein distance and the style of feature maps derived from the early layers of the generator and the target domain images (see figure \ref{Figure1}).

Our approach is inspired by methods \cite{gatys2016image,zhang2019noise}, demonstrating the independent processing of feature representations learned by neural networks. Our features transfer approach combines a domain-to-domain translation concept, which builds a mapping between two domains for adaptation instead of mapping between two individual images, unlike \cite{gatys2016image}, with a noise loss employing the Wasserstein distance and the stylistic features derived from the generator instead of the discriminator, unlike NAGAN \cite{zhang2019noise}.

We define the training source dataset as $S$ and the target dataset as $T$. Our main objective is to learn a mapping from $S$ to $T$ so that the feature distribution of image $x \epsilon S$ will be translated to $x^{'}$ images such that anatomical content is preserved while the style of image feature distribution is similar to that of $T$. The generator's job is to translate the source images that map into the target domain. The $D_n$ distinguishes the texture pattern and noise between domains $S$ and $T$. The $D_c$ distinguishes the anatomical content of images. $\mathcal{L}_{an}$ and $\mathcal{L}_{ac}$ are used for adversarial training. $L_c$ and $L_n$ are used to preserve the anatomical contents, transfer texture patterns and reduce the noise, respectively. Finally, adversarial losses are combined with content and noise losses to train the model. The computation of all loss functions is given below.

\subsubsection{Adversarial losses} The $\mathcal{L}_{an}$ and $\mathcal{L}_{ac}$ are used to train the $D_n$ and $D_c$, respectively. The $D_n$ job distinguishes between the target image $y \epsilon T$ and the translated image $G(x)$. The $D_c$ distinguishes between the translated image $G(x)$ and the real image $x \epsilon S$. The Computation of the losses are as follows: 

\begin{equation}
\mathcal{L}_{an}(G, D_n) = \mathbb{E}_{y \sim T} [\log D_n(y)] + \mathbb{E}_{x \sim S} [\log (1 - D_n(G(x)))]
\label{eq1}
\end{equation}

\begin{equation}
\mathcal{L}_{ac}(G, D_c) = \mathbb{E}_{x \sim S} [\log D_c(x)] + \mathbb{E}_{x \sim S} [\log (1 - D_c(G(x)))]
\label{eq2}
\end{equation}

\subsubsection{Content loss} To preserve the anatomical contents of the image, the content loss $L_c$ is computed from the last layer of generator $G$ as follows:

\begin{equation}
\mathcal{L}_c(G) = \mathbb{E}_{x \sim S} \left[ \left| F(G(x_i))^l - F(x_i)^l \right| \right]
\label{eq3}
\end{equation}

where $F(\cdot)^{l}$ represents the feature maps from the last layer.

\subsubsection{Noise loss} We formulate noise as style \cite{gatys2016image}. The style of an image was computed mathematically by Gram Matrix \cite{li2017demystifying} $G_r$ as follows: 

\begin{equation}
G_r(y)^{(G, l)}_{(i, j)} = \text{vec}\left[F(y)^{(G, l)}_{i}\right] \otimes \text{vec}\left[F(y)^{(G, l)}_{j}\right]
\label{eq4}
\end{equation}

where $G_{r}(y)^{(G,l)}$ is the Gram matrix of $l^{th}$ layer from generator's feature maps. The subscript $i$ and $j$ denotes the $i^{th}$ and $j^{th}$ maps of the $G_{r}(y)^{(G,l)}$, $vec[\cdot]$ represents the vectorization operation, and $\otimes$ represents the inner product.

$L_n$ is computed as follows: 

\begin{equation}
\mathcal{L}_n(G) = \mathbb{E}_{(x, y) \sim (S, T)} \left[ \sum_{l=1}^{3} W\left(G_r(G(x)), G_r(y)\right) \right]
\label{eq5}
\end{equation}

Where $W$ is the Wasserstein distance \cite{arjovsky2017wasserstein}, we trained our model with a feature map of various early layers such as 1, 2, and 3 and decided to keep 3 for final results based on the outcomes. Thus, we used only the first three layers of the feature map corresponding to noise. The $l$ denotes the layers in the generator.

\subsubsection{Objective function} Full objective function is calculated as follows:

\begin{equation}
\arg{\min\max}_{G, D_n, D_c} \mathcal{L}(G, D_n, D_c) = \mathcal{L}_{an}(G, D_n) + \mathcal{L}_{ac}(G, D_c) + \lambda_1 \cdot \mathcal{L}_c(G) + \lambda_2 \cdot \mathcal{L}_n(G)
\label{eq6}
\end{equation}

Where $\lambda_1$ and $\lambda_2$ are the hyperparameters, our aim is to optimize the above equation.

\subsection{Implementation Details}
\subsubsection{Network architecture} 
We adopted the generator and discriminator networks architecture from Jun-Yan Zhu et al. \cite{zhu2017unpaired}. The generator is a Resnet-15 network that we modified by adding one extra convolution block and one deconvolution block to manage the feature map extraction from the first three layers. The generator network contains nine residual blocks (including a reflection padding operation, an instance normalization layer, a 2-D convolution layer, a ReLu activation function, an instance normalization layer, a 2-D convolution layer, and a plus operation sequentially), four convolutional blocks (including a ReLu activation function, an instance normalization layer, and a 2-D convolution layer) and four deconvolutional blocks (including a ReLu activation function, an instance normalization layer, and a 2-D deconvolution layer). 
The structure of both discriminators is the same. It contains four convolutional blocks (including a ReLu activation function, an instance normalization layer, and a 2-D convolution layer) and a fully connected layer. 

\subsubsection{Training details}
We used Adam optimizer \cite{kingma2014adam} with batch size 1 to train the model to 200 epochs. The learning rate was kept at 0.0002 for the first 100 epochs, then linearly decreased to zero over the following 100 epochs. The $\lambda_1$ and $\lambda_2$ values in equation 6 were set to 10 for all experiments. The number of filters in the generator and discriminator’s first convolutional layer was 64. Model training was done using two NVIDIA GeForce RTX 2080TI GPUs.

\subsubsection{Datasets}
We retrospectively included carotid ultrasound 2D images from the longitudinal VIPVIZA study cohort of Umea University \cite{naslund2019visualization}. The study population are healthy subjects (50:50 men and women, 40-60 years old at base-line examination), with subclinical atherosclerosis. Expert operators acquired images in a longitudinal 2D B-mode of the carotid arteries. In the baseline and 3-year follow-up of the VIPVIZA study, a Cardio Health Station (CHS) and a linear 7 MHz transducer (Panasonic Healthcare Corporation of North America, Newark, NJ, USA) ultrasound machine (“US System A”) were used. In the 6-year follow-up of the study, a GE Vivid IQ with the 9 MHz transducer (GE HealthCare, Chicago, IL, USA) ultrasound machine (“US System B”) was used.

\begin{table*}[!ht]
\caption{Description of the images used in the experiments. Here, GSM, CHS, GE IQ, and VIPVIZA stand for grey scale median, cardio health station, general electric image quality, and västerbotten intervention program visualization of atherosclerosis, respectively.}
\begin{center}
\setlength{\tabcolsep}{2pt}
\begin{tabular}{|c|c|c|c|c|}
\hline
\rule{0pt}{1.0\normalbaselineskip}
\multirow{2}{*}{Characteristics} & \multicolumn{2}{|c|}{Experiment 1} & \multicolumn{2}{|c|}{Experiment 2} \\[5pt]
\cline{2-5}
\rule{0pt}{1.0\normalbaselineskip}
 &  Domain A & Domain B & Domain C & Domain D \\[5pt]
\hline
\newcommand\Tstrut{\rule{0pt}{2.6ex}}
\rule{0pt}{1.0\normalbaselineskip}
Anatomical content & Carotid plaques & Carotid plaques  & Carotid intima-media & Carotid intima-media \\[5pt]
Projection & Longitudinal & Longitudinal  & Longitudinal & Longitudinal \\[5pt]
US system & CHS (sys A) & GE IQ (sys B) & mixed & mixed \\[5pt]
Source & VIPVIZA 3y FU & VIPVIZA 6y FU & mixed & mixed \\[5pt]
Noise level & mixed & mixed & high & low \\[5pt]
N train & 443 & 546 & 225 & 215 \\[5pt]
N test & 50 & - & 15 & - \\[5pt]
Plaque GSM & 44.6 (22.4) & 48.6 (20.3) & - & - \\[5pt]
\hline
\end{tabular}
\end{center}
\label{tab1}
% Footer
\end{table*}

\paragraph{Experiment 1 – Image harmonization:}
In this experiment, we included carotid plaque images from the two different systems (System A and System B). The task for the domain adaptation was to adapt the system A image features to resemble those of system B images while retaining the anatomical content.
For system A, the plaque acquisition mode was used, producing a B-mode image with no harmonic imaging and spatial angular compounding of three projections – here defined as “domain A”. Images were reconstructed offline from the .cine format, and 493 consecutive images from different subjects were included from system A, and 546 for system B.
For system B, standard B-mode acquisition mode was used, including harmonic imaging, and images were stored in dicom screen capture format – here defined as “domain B”.
Images had 8-bit intensity depth, and all images were resampled to 400x400 pixels, corresponding to a spatial resolution on the order of 0.1mm/px.

\begin{figure*}
\centering
{\includegraphics[scale=0.60]{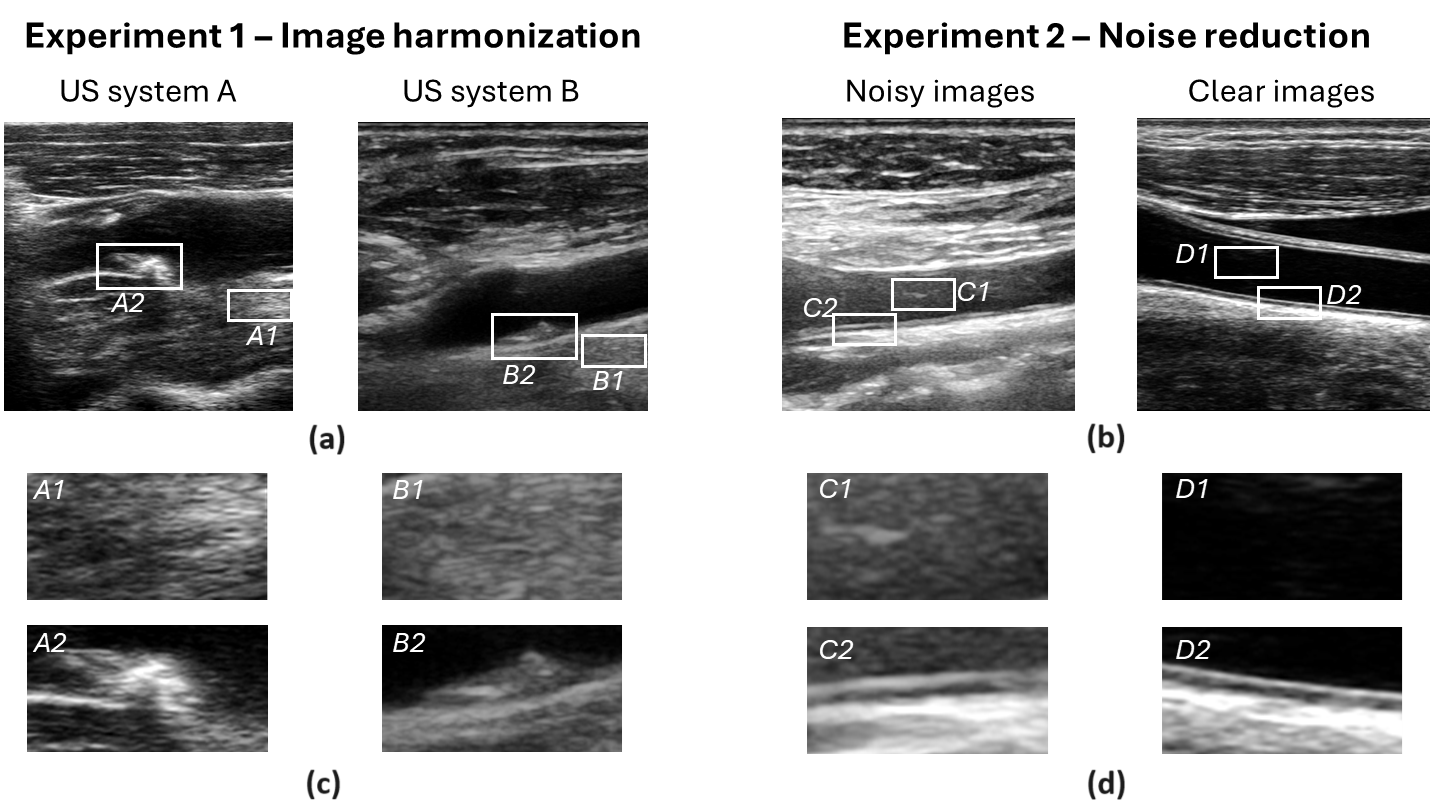}}
\caption{Examples of carotid ultrasound images used in the training of the domain adaptation task for the two experiments. \textbf{a)} show examples for experiment 1 (image harmonization) that we define as domains A and B, respectively. \textbf{b)} show examples for experiment 2 (noise reduction) that we define as domains C and D. In the lower panel, \textbf{c)} and \textbf{d)}, regions of interests (ROIs) from the upper images are shown to zoom in on a homogeneous tissue segment (A1 and B1), carotid plaques (A2 and B2), Lumen/blood (C1 and D1), and arterial walls (C2 and D2).}
\label{Figure2}
\end{figure*}

\paragraph{Experiment 2 – Noise reduction:}
In this experiment, we included carotid images of the intima-media complex (without plaques) from both systems A and B. 
For system A, the CHS IMT mode, which uses harmonic imaging for intima-media images and measurement, was used. For system B, conventional B-mode and harmonic imaging were used, and images were acquired to measure the intima-media thickness.
Out of these images, noisy and clear images were manually classified by an expert operator (biomedical scientist and sonographer with +10 years of clinical ultrasound) - noisy and clear images were defined as “domain C” and “domain D”, respectively.
Noisy images were included based on the criteria of visual noise in the lumen of the images (a haze of bright pixels in the region of the blood), and clear images as its counterpart – with no noise in the lumen (dark pixels in the region of the blood). See figure \ref{Figure2} for examples. 
The task for the domain adaptation was to adapt the noisy image features to those of the clear images while retaining the anatomical content,
Images had 8-bit intensity depth, and all images were resampled to 400x400 pixels, corresponding to a spatial resolution on the order of 0.1mm/px.

\subsection{Evaluation}
The performance of the translation was evaluated from four perspectives: 1) their feature distribution similarity, 2) pixel-space similarity, 3) down-stream estimation of an image-based cardiovascular risk marker, and 4) (reverberation) noise level. For comparison, we also trained the default cycleGAN model.

The similarity in feature distributions was computed both without any translation and with translation, using the Bhattacharya distance (BD) and Histogram correlation (HC) \cite{bhattacharyya1943measure} metrics. Mathematically, they are given by:

\begin{equation}
BD(h_1, h_2) = \sqrt{1 - \frac{1}{\sqrt{\bar{h}_1 \bar{h}_2 N^2}} \sum_I \sqrt{h_1(I) h_2(I)}}
\label{eq8}
\end{equation}

\begin{equation}
HC(h_1, h_2) = \frac{\sum_I (h_1(I) - \bar{h}_1) (h_2(I) - \bar{h}_2)}{\sqrt{\sum_I (h_1(I) - \bar{h}_1)^2 \sum_I (h_2(I) - \bar{h}_2)^2}}
\label{eq7}
\end{equation}

Where $h_1$ and $h_2$ are the normalized histograms of the images. $\bar{h}_1$ and $\bar{h}_2$ represents the mean value of the histograms $h_1$ and $h_2$, respectively. $N$ is the total number of bins in the histogram. $I$ represent the background region pixels. High HC and low BD between images represent high feature distribution similarity. 

The pixel-space similarity was computed using the structure similarity index measure (SSIM) between the input image and the corresponding translated image. The purpose was to evaluate the preservation of anatomical content. The SSIM metric is based on different regional pixel statistics in the image, and a high value means strong similarity. The SSIM was computed for the whole image and individual tissues by averaging manually segmented regions of interest (ROIs) of the plaque/intima-media tissues, the lumen, and adventitia (e.g. figures \ref{Figure5} and \ref{Figure7}).

To assess the impact of the translation on image-based risk markers, we computed the grey scale median (GSM) and plaque vulnerability re-classification rate. The GSM is a risk marker that is related to plaque vulnerability \cite{nicolaides2010asymptomatic}, where a low value is related to symptomatic (vulnerable) plaques and a high value to asymptomatic (stable) plaques. One threshold that has been used is GSM=25 \cite{christodoulou2003texture}. Using this threshold, we computed the number of plaques that were re-classified, i.e. crossed this threshold due to the domain adaptation.

\begin{equation}
GSM = \text{median} \left( 190 \cdot \left( I_{\text{whole\_image}} - \text{mean}(I_{\text{ROI\_Lumen}} \right) \right) / \max(I_{\text{ROI\_Adventitia}}) 
\label{eq9}
\end{equation}

Finally, the noise level was estimated using the lumen and wall ROIs as

\begin{equation}
\text{Contrast} = -20 \log_{10} \left( \frac{\text{mean}(I_{\text{ROI\_Lumen}})}{\text{mean}(I_{\text{ROI\_Adventitia}})} \right)
\label{eq10}
\end{equation}

If there is low noise, the lumen should appear dark, and the $I_{\text{ROI\_lumen}}$ should have low values, causing the Contrast to be strongly negative in the dB scale.

\section{Results}

\subsection{Experiment 1 – Image harmonization}
\begin{figure*}
\centering
{\includegraphics[scale=0.60]{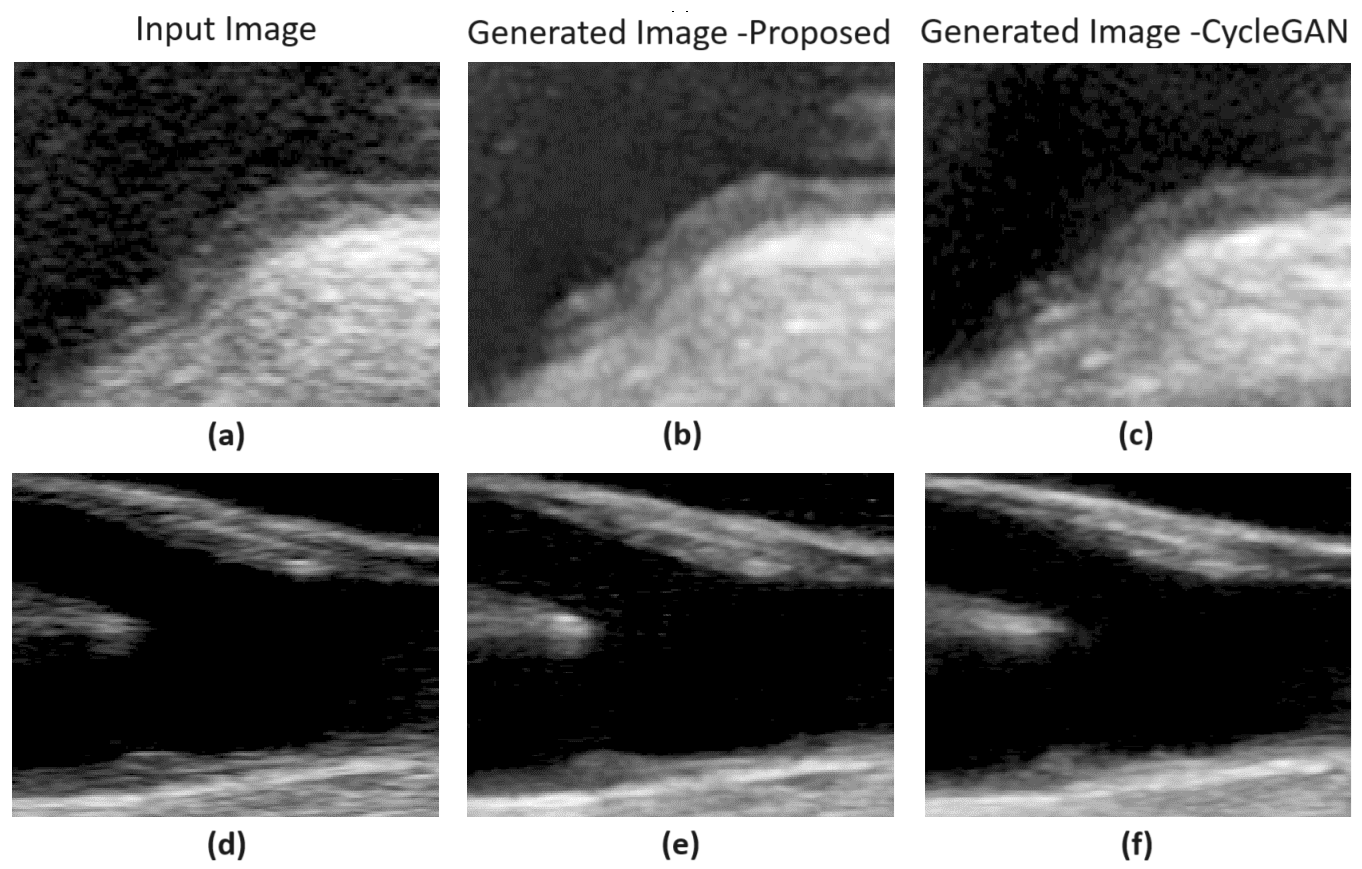}}
\caption{Example of results for experiment 1 (Image harmonization). a) and d) shows input images (Domain A). b) and e) shows the proposed model's generated images. c) and f) shows the corresponding generated images of the cycleGAN model.}
\label{Figure3}
\end{figure*}

\begin{figure*}
\centering
{\includegraphics[scale=0.70]{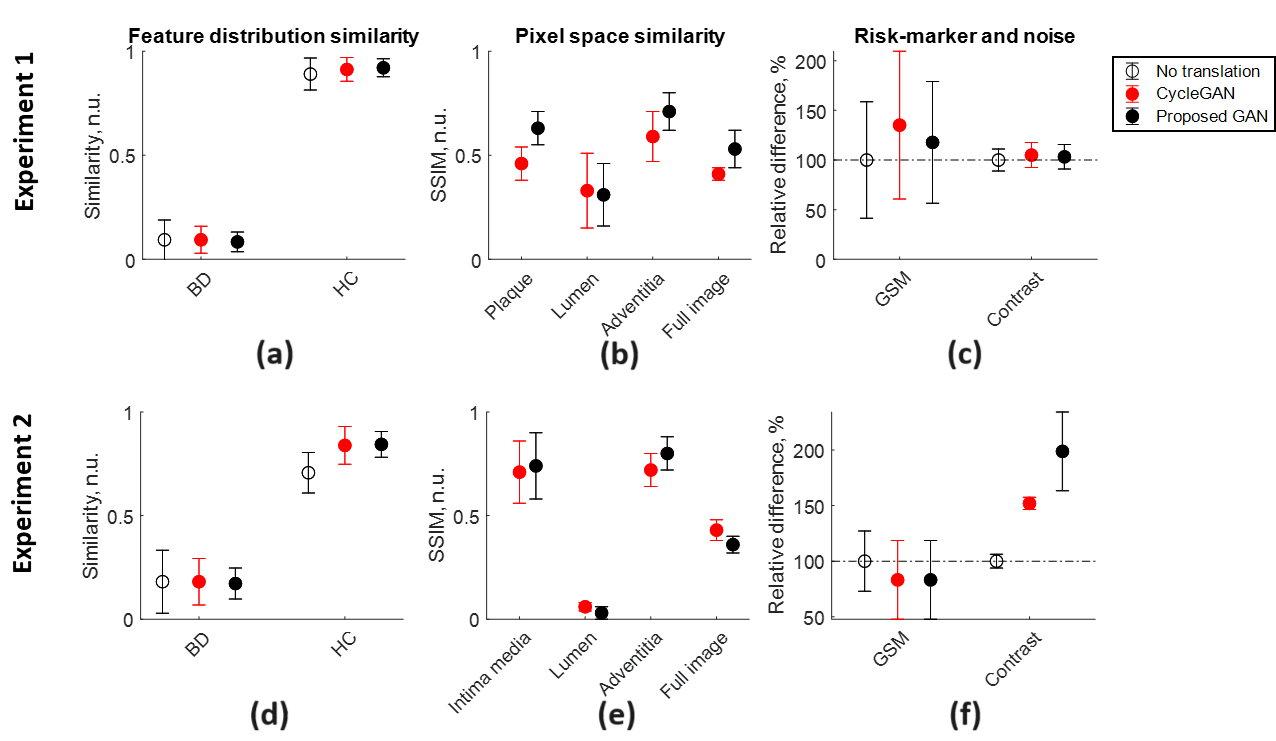}}
\caption{Summary of results. The upper panel is for experiment 1 (image harmonization), and the lower panel is for experiment 2 (noise reduction). The evaluation was done in four perspectives: 1) feature distributions, \textbf{a)} and \textbf{d)}, 2) pixel space \textbf{b)} and \textbf{e)}, and 3) image-based risk marker for cardiovascular disease (GSM) and 4) noise level (Contrast), \textbf{c)} and \textbf{f)}. Feature distributions were compared using bhattacharyya distance (BD) and histogram correlation (HC), and pixel-space similarity was evaluated using the structure similarity index measure (SSIM). The risk marker and contrast measures are presented as the relative change between the corresponding input and domain-adapted output images.}
\label{Figure4}
\end{figure*}

Figure \ref{Figure3} shows examples of two domain A images (cropped) and their corresponding translations using cycleGAN and the proposed GAN. It can be seen that the anatomical content of the input images is retained in the translated versions, but the fine-detailed texture pattern is modified. Evaluation of feature distribution similarity showed that both translation models improved the similarity, but the proposed model had the highest performance (Table \ref{tab2} and Figure \ref{Figure4}). The proposed model achieved a lower BD and higher HC compared to CycleGAN and no translation, demonstrating that our model outperformed CycleGAN. 

\begin{figure*}
\centering
{\includegraphics[scale=0.55]{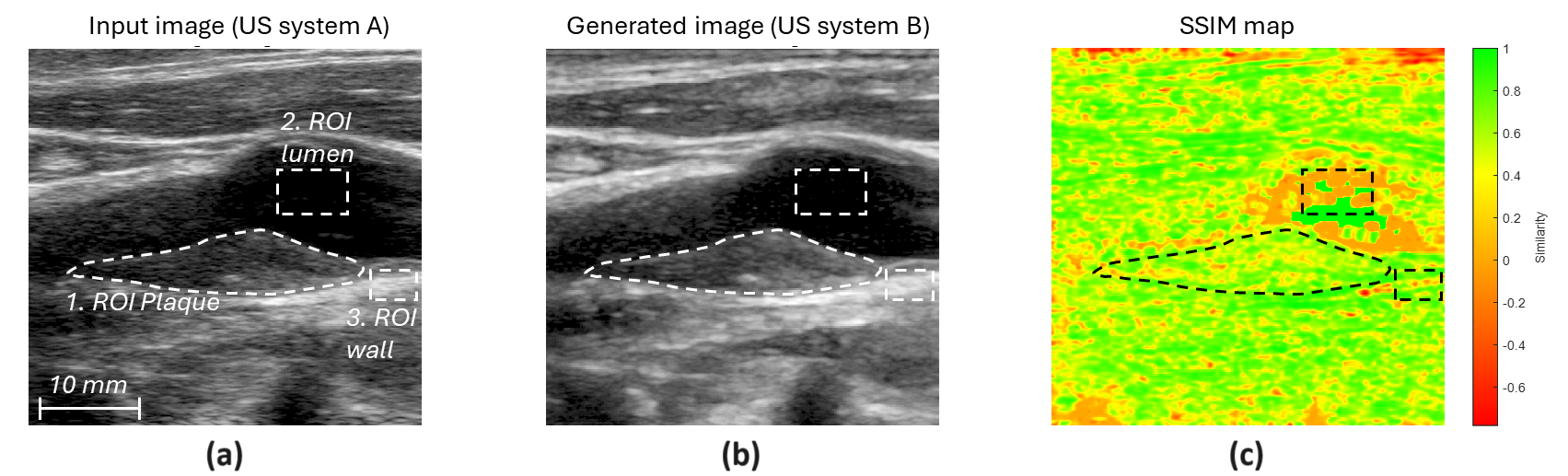}}
\caption{Illustration of ROIs used in Experiment 1 to assess translation performance (Lumen, Plaque and Wall/Adventitia). \textbf{a)} shows an example image of domain A, \textbf{b)} the generated image (domain B) and \textbf{c)} the corresponding structure similarity index measure (SSIM) map.}
\label{Figure5}
\end{figure*}

Figure \ref{Figure5} shows an example of an input image, generated translation and the corresponding pixel-space similarity comparison using SSIM. Evaluation of pixel-space similarity showed that SSIM had higher values (higher similarity) for the proposed model than the GAN model in all tissues except the lumen (Table \ref{tab3} and Figure \ref{Figure4}).

The difference in plaque GSM between input and output images was 7.6 (6.5) vs 15.0 (13.4) for the proposed and cycleGAN models, respectively (Table \ref{tab4}). In the training data, the difference in GSM plaque between domains B and A was 4.0 (p<0.006, independent samples t-test).
The contrast between the arterial wall (adventitia) and lumen ROIs was -34.1 (3.8) dB, -35.2 (4.1) dB and -35.8 (4.3) dB for the original image, cycleGAN, and proposed model translations (Table \ref{tab4}).

\begin{table*}[!ht]
\caption{Evaluation of feature transfer for Experiment 1 (Image Harmonization). To measure the dissimilarity and correlation between two domain images, bhattacharyya distance (BD) and histogram correlation (HC) are used. The results include mean and standard deviation (SD) values of these metrics.}
\begin{center}
\setlength{\tabcolsep}{5pt}
\begin{tabular}{|c|c|c|}
\hline
\rule{0pt}{1.0\normalbaselineskip}
\multirow{2}{*}{Method} & \multicolumn{2}{|c|}{Mean(SD)} \\[5pt]
\cline{2-3}
\rule{0pt}{1.0\normalbaselineskip}
 &  BD & HC \\[5pt]
\hline
\newcommand\Tstrut{\rule{0pt}{2.6ex}}
\rule{0pt}{1.0\normalbaselineskip}
No Translation A vs A & 0.044 (0.024) & 0.958 (0.022) \\[5pt]
No Translation B vs B & 0.033 (0.012) & 0.968 (0.011) \\[5pt]
No Translation A vs B & 0.120 (0.095) & 0.890 (0.077) \\[5pt]
CycleGAN A to B' & 0.094 (0.065) & 0.912 (0.057) \\[5pt]
Proposed model A to B' & 0.084 (0.047) & 0.920 (0.043) \\[5pt]
\hline
\end{tabular}
\end{center}
\label{tab2}
\end{table*}

\begin{table*}[!ht]
\caption{Performance evaluation for pixel-space similarity for whole image and individual tissues - Experiment 1 (Image harmonization).}
\begin{center}
\setlength{\tabcolsep}{5pt}
\begin{tabular}{|c|c|c|c|c|}
\hline
\rule{0pt}{1.0\normalbaselineskip}
\multirow{2}{*}{Model} & \multicolumn{4}{|c|}{Tissue segment} \\[5pt]
\cline{2-5}
\rule{0pt}{1.0\normalbaselineskip}
 &  Plaque & Lumen & Adventitia & Whole Image \\[5pt]
\hline
\newcommand\Tstrut{\rule{0pt}{2.6ex}}
\rule{0pt}{1.0\normalbaselineskip}
Proposed GAN & 0.63 (0.08) & 0.31 (0.15) & 0.71 (0.09) & 0.53 (0.09) \\[5pt]
CycleGAN & 0.46 (0.08) & 0.33 (0.18) & 0.59 (0.12) & 0.41 (0.03) \\[5pt]
\hline
\end{tabular}
\end{center}
\label{tab3}
\end{table*}

\begin{table*}[!ht]
\caption{Performance evaluation - Impact on risk markers and blood-to-tissue contrast - Experiment 1 (Image harmonization). GSM stands for grey scale median.}
\begin{center}
\setlength{\tabcolsep}{5pt}
\begin{tabular}{|c|c|c|c|c|}
\hline
\rule{0pt}{1.0\normalbaselineskip}
 &  A & B' & Diff & p \\[5pt]
\hline
\rule{0pt}{1.0\normalbaselineskip}
&  GSM &  &  &  \\[5pt]
\newcommand\Tstrut{\rule{0pt}{2.6ex}}
\rule{0pt}{1.0\normalbaselineskip}
Proposed GAN & 42.7 (25.0) & 50.3 (26.2) & 7.6 (6.5) & P<0.001 *** \\[5pt]
CycleGAN  & 42.7 (25.0)  & 57.7 (31.8) & 15.0 (13.4) & P<0.001 *** \\[5pt]
          & Re-classification &            &         &             \\[5pt]
Proposed GAN & 5 (10\%)  &      &     &  \\[5pt]
CycleGAN &   12 (24\%)   &   &    &      \\[5pt]
        & Contrast (dB)  &   &   &          \\[5pt]
Proposed GAN &  -34.1 (3.8)	& -35.2 (4.1) &  &  P=0.78    \\[5pt]
CycleGAN & 	-34.1 (3.8) & -35.8 (4.3) &	 & P=0.73         \\[5pt]
\hline
\end{tabular}
\end{center}
\label{tab4}
% Footer
\hspace{2.5cm} \textit{Plaque area = 24.9 (18.0) mm$^2$}
\end{table*}

\subsection{Experiment 2 – Noise reduction}
\begin{figure*}
\centering
{\includegraphics[scale=0.60]{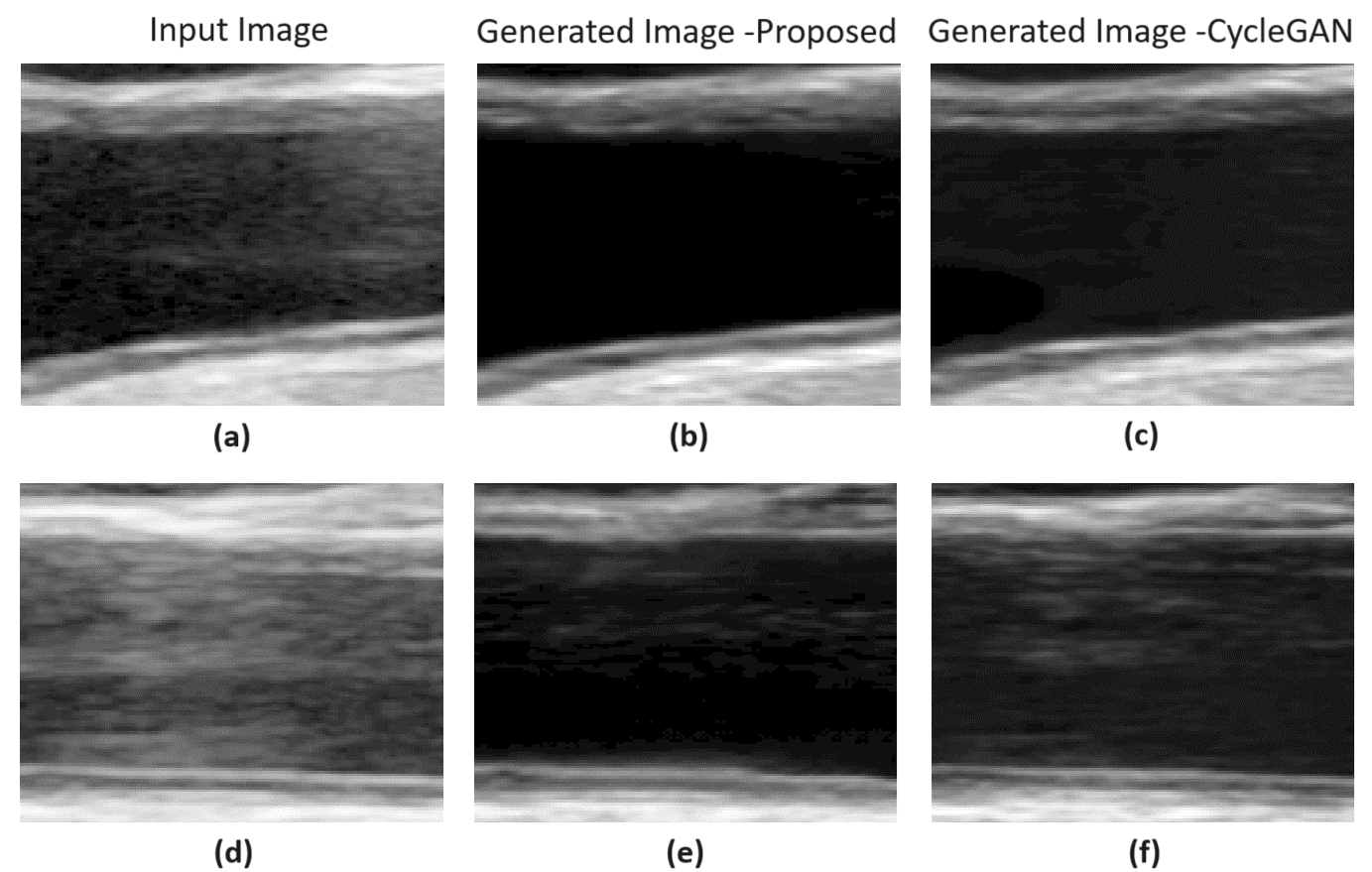}}
\caption{Example of results for experiment 2 (Noise reduction). \textbf{a)} and \textbf{d)} shows input images (Domain C) and \textbf{b)} and \textbf{e)} shows the proposed model's generated images. \textbf{c)} and \textbf{f)} show the corresponding generated images of the cycleGAN model.}
\label{Figure6}
\end{figure*}

Figure \ref{Figure6} shows examples of two noisy domain C images (cropped) and their corresponding translations using cycleGAN and the proposed GAN, where noise in the lumen region was reduced.

\begin{figure*}
\centering
{\includegraphics[scale=0.55]{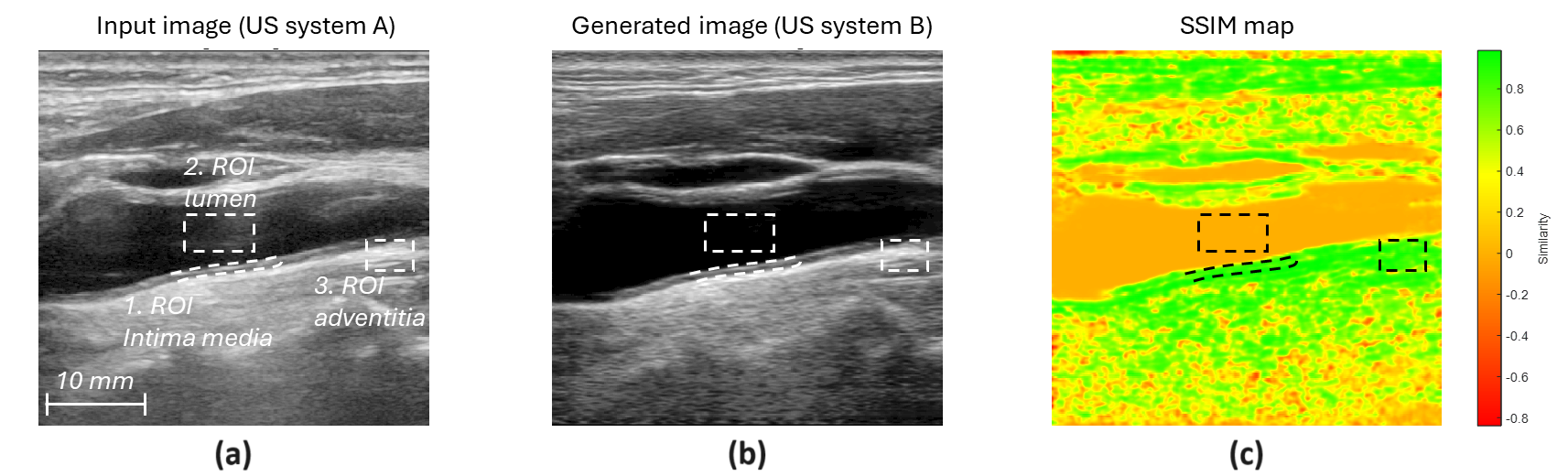}}
\caption{Illustration of ROIs used in Experiment 2 to assess translation performance (Lumen, inner arterial wall (intima media) and outer wall (adventitia)). \textbf{a)} shows an example image of domain C, \textbf{b)} the generated image (domain D) and \textbf{c)} the corresponding structure similarity index measure (SSIM) map.}
\label{Figure7}
\end{figure*}

The feature similarity distribution metrics between input and translated images showed that both models improved the BD and HC similarities as compared to no translation and that the proposed model was the best (Table \ref{tab5}, Figure \ref{Figure4}). 
Figure \ref{Figure7} shows example results for noise suppression in carotid ultrasound images by SSIM map and illustrates that the translation models strongly modify the lumen ROI while retaining the overall anatomy. The pixel-space similarity estimations (SSIM) in different tissues also showed that the lumen was strongly dissimilar with SSIM values close to 0, whereas tissues had strong similarity ~0.7-0.8 for both models.

Evaluation of the influence of translation on risk marker showed that the intima-media GSM changed from X to Y and Z for the cycleGAN and proposed GAN models, respectively.
The contrast between adventitia and lumen ROIs changed from -23.5 (3.2) dB to -35.7 (2.8) dB and -46.7 (18.1) dB for the input image and the cycleGAN and proposed GAN translations, respectively. Note: The proposed model had the most significant contrast, indicating the highest noise reduction.

\begin{table*}[!ht]
\caption{Evaluation of feature transfer for Experiment 2 (Noise Reduction). To measure the dissimilarity and correlation between two domain images, bhattacharyya distance (BD) and histogram correlation (HC) are used. The results include mean and standard deviation (SD) values of these metrics.}
\begin{center}
\setlength{\tabcolsep}{5pt}
\begin{tabular}{|c|c|c|}
\hline
\rule{0pt}{1.0\normalbaselineskip}
\multirow{2}{*}{Method} & \multicolumn{2}{|c|}{Mean(SD)} \\[5pt]
\cline{2-3}
\rule{0pt}{1.0\normalbaselineskip}
 &  BD & HC \\[5pt]
\hline
\newcommand\Tstrut{\rule{0pt}{2.6ex}}
\rule{0pt}{1.0\normalbaselineskip}
No Translation C vs C & 0.024 (0.008) & 0.976 (0.007) \\[5pt]
No Translation D vs D & 0.015 (0.008) & 0.985 (0.007) \\[5pt]
No Translation C vs D & 0.357 (0.152) & 0.707 (0.098) \\[5pt]
CycleGAN C to D'      & 0.181 (0.112) & 0.839 (0.091) \\[5pt]
Proposed model C to D' & 0.172 (0.075) & 0.844 (0.062) \\[5pt]
\hline
\end{tabular}
\end{center}
\label{tab5}
\end{table*}

\begin{table*}[!ht]
\caption{Performance evaluation for pixel-space similarity for whole image and individual tissues - Experiment 2 (Noise reduction).}
\begin{center}
\setlength{\tabcolsep}{5pt}
\begin{tabular}{|c|c|c|c|c|}
\hline
\rule{0pt}{1.0\normalbaselineskip}
\multirow{2}{*}{Model} & \multicolumn{4}{|c|}{Tissue segment} \\[5pt]
\cline{2-5}
\rule{0pt}{1.0\normalbaselineskip}
 &  IM & Lumen & Adventitia & Whole Image \\[5pt]
\hline
\newcommand\Tstrut{\rule{0pt}{2.6ex}}
\rule{0pt}{1.0\normalbaselineskip}
Proposed GAN & 0.74 (0.16) & 0.03 (0.03) & 0.80 (0.08) & 0.36 (0.04)\\[5pt]
CycleGAN & 0.71 (0.15) & 0.06 (0.02) & 0.72 (0.08) & 0.43 (0.05) \\[5pt]
\hline
\end{tabular}
\end{center}
\label{tab6}
\end{table*}

\begin{table*}[!ht]
\caption{Performance evaluation - Impact on riskmarkers and blood-to-tissue contrast - Experiment 2 (Noise reduction). GSM stands for grey scale median.}
\begin{center}
\setlength{\tabcolsep}{5pt}
\begin{tabular}{|c|c|c|c|c|}
\hline
\rule{0pt}{1.0\normalbaselineskip}
Method &  C & D' & Diff & p \\[5pt]
\hline
\rule{0pt}{1.0\normalbaselineskip}
&  GSM &  &  &  \\[5pt]
\newcommand\Tstrut{\rule{0pt}{2.6ex}}
\rule{0pt}{1.0\normalbaselineskip}
Proposed GAN & 51.2 (13.9) & 45.8 (17.7) & -5.3 (11.8) & P=0.10 \\[5pt]
CycleGAN  & 51.2 (13.9) & 42.6 (18.1) & -8.6 (6.2) & P<0.001 *** \\[5pt]
        & Contrast (dB)  &   &   &          \\[5pt]
Proposed GAN &  -23.5 (3.2) & -46.7 (18.1) &   & P<0.001 ***    \\[5pt]
CycleGAN     & 	-23.5 (3.2) & -35.7 (2.75) &   &  P<0.001 ***   \\[5pt]
\hline
\end{tabular}
\end{center}
\label{tab7}
\end{table*}

\section{Discussion}
In this work, we proposed a GAN-based model for multi-purpose domain adaptation tasks and tested its performance for image harmonization between carotid ultrasound images captured with different ultrasound systems and image noise reduction in carotid ultrasound images with high noise. The main results show that 1) the model can adapt the image features between two different ultrasound systems, as well as reduce the noise in noisy images while still retaining the overall anatomy (content) in the images, and 2) the domain adaptation influenced critical risk markers for cardiovascular risk computed on the input and output images.

\subsection{Image harmonization}
Results showed adapted feature distributions compared to those without adaptation (0.12 vs 0.89) but not as similar to those within domains (0.03 vs 0.97) for BD and HC, respectively (Table \ref{tab2}).
The feature similarities were higher than that of the NAGAN model \cite{zhang2019noise}, where they found 0.32 and 0.58 for BD and HC and 0.36 and 0.58 for the CycleGAN model. However, it should be noted that they adapted OCT and US images and, therefore, have larger differences between the image domains.
The underlying content (anatomy) of the images was retained, as shown by the SSIM values (Table \ref{tab3}). For the whole image, SSIM was 0.53 (0.09). This is lower than those found by Ali et al. \cite{ali2023translation}, translating US images of healthy and diseased arteries, with an overall SSIM of 0.78 (0.02). However, those images only included the common carotid artery, where the anatomy has less variation than the bulb. The bulb is the position where the carotid artery divides into two arteries, the internal and external carotids. The latter was often included in the images of this study since this is where the plaques most commonly form. In addition, it should be noted that the SSIM were higher in our results for plaque and adventitia arterial wall segments SSIM were 0.63 and 0.71 as compared to the overall image - and these tissues are the ones that are used for assessing cardiovascular disease \cite{nicolaides2010asymptomatic}. Thus, a lower preservation of the overall details of the image anatomy is likely not critical for our application.

\subsection{Noise reduction}
The primary source of noise in the ultrasound images is caused by diffuse reverberation which is caused by the multiple reflections of the transmitted ultrasound pulse in the tissue before returning to the transducer.
The results of this work showed adapted feature distributions compared to those without adaptation (0.36 vs 0.71) but not as similar to those within domains (0.02 vs 0.98) for BD and HC, respectively (Table \ref{tab5}).
The noise reduction in the image can be clearly seen in Figure \ref{Figure6} and is supported by the low SSIM values in the lumen (0.03), while the arterial wall tissues are unaffected (SSIM 0.74 and 0.80), as shown in Table \ref{tab6}). 
In addition, the noise reduction was significant for both models, but the proposed model was better, as seen in the Contrast measurements changing from -23.5 dB to -46.7 dB and -35.7 dB for the proposed vs CycleGAN models, respectively (Table \ref{tab7}). Recent work by other groups, e.g., by Brickson et al. \cite{brickson2021reverberation}, used a 3D CNN model on the underlying radio-frequency data of ultrasound images and showed a Contrast reduction from -18.9 (5.8) to -34.9 (8.1) dB. Thus, noise reduction performance in our work is similar and potentially better. 
The proposed GAN preserved the wall and adventitia tissues the best, as seen in SSIM for wall and adventitia (Table \ref{tab6}). It should be noted that it slightly distorted other parts of the image (see Figure \ref{Figure7}), e.g., muscle and vein, which is also indicated by a slightly lower overall SSIM compared to experiment 1.
However, this may not be a critical problem for our application of assessing cardiovascular disease (as commented for experiment 1).

\subsection{Influence on image-based computation of cardiovascular risk markers}
In experiment 1, the impact of the domain adaptations on image-based risk markers showed that both models increased the GSM of the plaques, with a consequence of re-classification of 12 plaques (24\%) for cycleGAN and 5 (10\%) plaques for the proposed model. Thus, the proposed model preserves the computed risk marker integrity better. The SSIM was different in the lumen, adventitia, and plaque tissues, indicating that the model translated them differently (Eq \ref{eq9}). Since the GSM risk marker is computed from ROIs in the lumen, adventitia and arterial plaque, the former finding can likely explain this change in GSM. 
In comparison, our group previously found that the blood pressure variation over the cardiac cycle caused a 16\% re-classification in the same study population \cite{nyman2018risk}, and 30\% re-classification in another study population with asymptomatic atherosclerosis (more severe disease) \cite{stenudd2020ultrasound}. This indicates that the re-classification by our model is similar to or lower than other known modulators of GSM.

In contrast, in experiment 2, the risk marker IM-GSM was used to assess the impact on clinical assessment, and it showed a slight decrease in the IM-GSM for both models but with large variability. The IM-GSM risk marker has not been as extensively studied in the literature as the GSM for plaques, and its relevance for assessing cardiovascular risk remains to be established. However, IM-GSM has been shown to be associated with the progression of atherosclerotic disease in the same study population \cite{nyman2023carotid} and to be associated with plaque GSM in populations with more advanced disease \cite{ibrahimi2015common}. Since the GSM computations in experiments 1 and 2 use the same tissue ROIs (Lumen and adventitial wall), and since the lumen was much more modified in experiment 2 compared to experiment 1, GSM estimations will be influenced differently in the two experiments.
It should be noted that the increase in GSM of the plaques may be built into the models because the training target domain B had significantly higher GSM values of the plaques than source domain A (48 vs 44, see Table \ref{tab1}). However, based on these results, we propose that evaluation of influence on risk marker estimations should always be done and possibly adjusted for when applying domain adaptation models to carotid US images since our results show that different models gave different changes in the risk marker.

\subsection{The proposed model}
The proposed model addressed both image harmonization and noise reduction, and results show that it performed better than the state-of-the-art model, cycleGAN, in both tasks.
The proposed model was inspired by the methods \cite{gatys2016image,zhang2019noise}, demonstrating the independent processing of feature representations learned by neural networks. In our model, we leverage the potential of Wasserstein distance to compute the noise loss using the style of feature maps driven from the early layers of the generator and the target domain images. The Wasserstein distance reduces dissimilarity by finding the most efficient way to map the distribution of one image set to another and accounting for the spatial relationships within the images by considering the geometry of the data in images\cite{arjovsky2017wasserstein, dukler2019wasserstein}. We consider that this makes it more suitable for comparing images where the spatial relationship between pixels is crucial, such as in medical images (e.g., carotid ultrasound images). Our features transfer approach builds a domain-to-domain mapping for both image harmonization and noise reduction.

We conducted experiments with both Resnet and Unet architectures as generators and observed that Resnet outperformed Unet in our applications. Therefore, the final results are reported based on the use of Resnet. Our generator employs a ResNet-15 architecture, while the discriminator is constructed using a classical CNN. The ResNet holds an advantage over CNN due to its incorporation of residual units, enabling deeper layers that directly leverage information from shallower ones. Consequently, feature maps extracted from the ResNet-15 are likely to exhibit more refined patterns and smoother transitions compared to those from a standard CNN. In addition, we fine-tuned the generator architecture from Jun-Yan Zhu et al. \cite{zhu2017unpaired} by adding two extra blocks (one convolution block and one deconvolution block) to manage the feature map extraction from the first three layers. We trained our model for various layers and decided to use the first three layers of the generator for stylistic feature extraction based on the training outcomes.

\subsection{Limitations and future work}
The proposed GAN model is designed for unsupervised image-to-image translation between two specific domains. It may not generalize well to tasks involving other domains or tasks with highly diverse datasets.

The features of carotid ultrasound images are influenced both by the specific imaging system (producing e.g. different spatial resolution and noise levels) as well as the severity of the disease (resulting in changes in the arterial wall, including plaques of various sizes and composition). While the present work and previous work by others show that adaptation between two different imaging systems can be done, the population is subclinical, meaning that the amount of disease is relatively low, and it is not clear how different amounts of disease in the arteries affect the translation performance. A potential solution to this could be to, e.g. train the model with a dataset including a larger distribution of the amount of disease in the arteries.

The results of the image harmonization task are between two specific ultrasound systems and will not, by default, generalize to other US systems. For other pairs of US systems, the model needs to be retained.
The noise reduction task was trained using images from both systems A and B, potentially allowing the model to learn more robust features (i.e., features not related to a specific system). Therefore, this model could potentially generalize to some degree to images of other ultrasound systems. The next step should be to train it on a larger set of images with low and high image quality and from multiple systems and use an external cohort for validation.

Moreover, in future work, the models could be trained using the risk marker data as input to the discriminator to minimize the change in risk markers that we found in this work.

The implementation of the model is a bit complex compared to traditional GAN architectures because it involves modifying the generator network to adjust harmonization features and noise levels between domains adaptively. This increased complexity can lead to higher computational overhead during training and inference. 

\section{Conclusion}
This paper proposes a GAN-based domain adaptation model for image harmonization and noise reduction of carotid ultrasound images. The results show that the model can 1) adapt features of ultrasound images of one system to images from another and 2) reduce the noise in the lumen (blood), both without changing the anatomical contents of the images. The proposed model performed better than the state-of-the-art model, CycleGAN. In addition, the results showed that the downstream task of computing image-based risk markers for cardiovascular risk assessment was, in general, influenced by the domain adaptations. We conclude that domain adaptation is a powerful tool to improve image quality and harmonization of image features in carotid ultrasound images, but the influence on computed risk markers should be evaluated for the specific model used.

\paragraph{Acknowledgment:}

This work was supported by the Heart Foundation of Northern Sweden, the Kempe foundations (JCK-3172), and Region Västerbotten. 

%Bibliography
\bibliographystyle{unsrt}  
\bibliography{arxiv}

\end{document}